\documentclass{sigchi}

\usepackage{balance}  
\usepackage{graphics} 
\usepackage{graphicx}
\usepackage{pdfpages}
\graphicspath{{figures/}}
\usepackage[T1]{fontenc}
\usepackage{txfonts}
\usepackage{mathptmx}
\usepackage[pdftex]{hyperref}
\usepackage{color}
\usepackage{booktabs}
\usepackage{textcomp}
\usepackage{microtype} 

\usepackage{todonotes}

\def\plaintitle{Are we on the same learning curve: Visualization of Semantic Similarity of Course Objectives}

\def\emptyauthor{}
\def\plainkeywords{Data extraction; Natural Language Processing; Visualization}

\makeatletter
\def\url@leostyle{%
  \@ifundefined{selectfont}{
    \def\UrlFont{\sf}
  }{
    \def\UrlFont{\small\bf\ttfamily}
  }}
\makeatother
\def\pprw{8.5in}
\def\pprh{11in}

\definecolor{linkColor}{RGB}{6,125,233}
\hypersetup{%
  pdftitle={\plaintitle},
  pdfauthor={\emptyauthor},
  pdfkeywords={\plainkeywords},
  bookmarksnumbered,
  pdfstartview={FitH},
  colorlinks,
  citecolor=black,
  filecolor=black,
  linkcolor=black,
  urlcolor=linkColor,
  breaklinks=true,
}
\begin{document}
\title{\plaintitle}
\author{%
Atish Pawar\thanks{Corresponding author - apawar1@lakeheadu.ca.}, Sahib Budhiraja, Daniel Kivi, Vijay Mago\\
\affaddr{Department of Computer Science}\\
\affaddr{Lakehead University, Thunder Bay, Canada}\\
\email{$\{\text{apawar1,sbudhira,dkkivi,vmago}\}$@lakeheadu.ca}\\
}

\maketitle

\begin{abstract}
The course description provided by instructors is an important piece of information as it defines what is expected from the instructor and what he/she is going to deliver during a particular course. One of the key components of a course description is the \textit{Learning Outcomes} section.  The contents of this section are used by program managers who are tasked to compare and match two different courses during the development of Transfer Agreements between different institutions. This research introduces the development of visual tools for understanding the two different courses and making comparisons.  We designed methods to extract the text from a course description document, developed an algorithm to perform semantic analysis, and displayed the results in a web interface.  We are able to achieve the intermediate results for the research which includes extracting, analyzing and visualizing the data.
\end{abstract}
\category{H.5.1. Computing methodologies}{}{}
\keywords{\plainkeywords}
\section{Introduction}
Learning Outcomes (LO) of a course defines what students are expected to learn by taking that course. Comparing the outcomes from two documents is a practice followed by program managers when they are asked to compare two courses or program \cite{CC:CC16}, for developing Transfer Program agreements between institutes. For instance, to give credits to a transfer student for a Programming Languages course being taught in a college, the process requires human intelligence and expertise to evaluate the course outcomes being followed by instructor at a college. For standardization, instructors are usually asked to follow Bloom's Taxonomy for developing LOs. Bloom's Taxonomy provides a general guidelines of keywords and a hierarchical structure to be used when defining the learning outcomes \cite{krathwohl2002}, see Figure 1\cite{forehand2010bloom}. But in practice, we found that this is what instructors usually don't do.

\begin{figure}[htbp]
\begin{center}
\includegraphics[width=0.4\textwidth, scale =0.5]{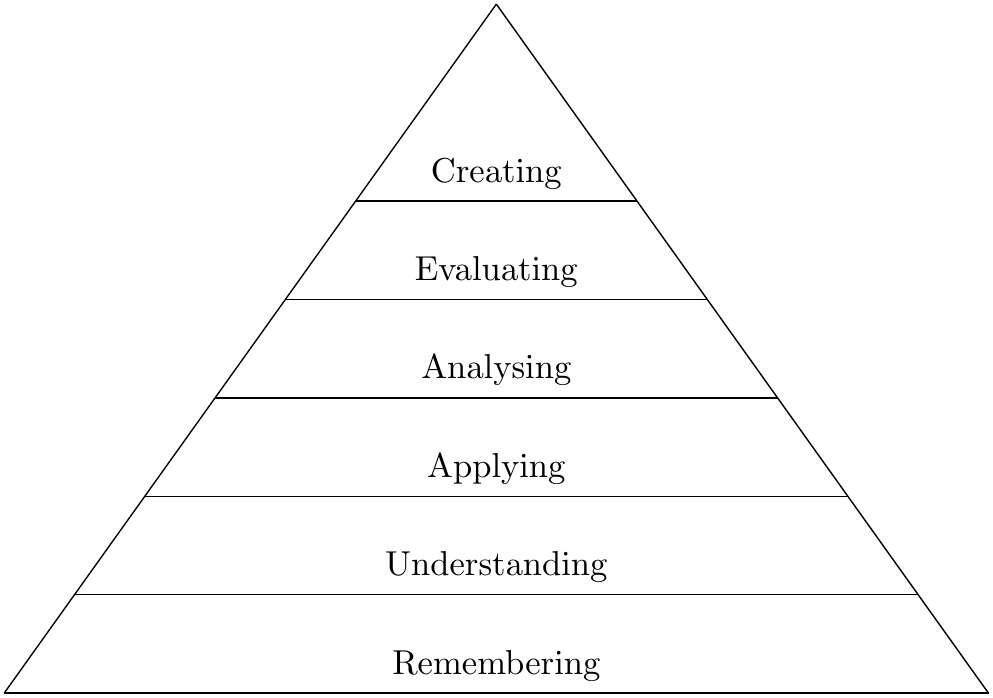}\\
\caption{Hierarchical Structure of Bloom's Taxonomy}
\label{Fig 1}
\end{center}
\end{figure}

The next challenge is extraction of LOs from the course description. This is tricky as all instructors are free to choose their own format and also there doesn't exist any specific standard format. Due to this, no two course outlines are the structured in the same way. Apart from that, the choice of words to define the LOs varies drastically. 

This research is aimed at automating the credit transfer process which currently requires human intelligence. The process currently involves a panel of people reviewing the course outlines and sometimes also requires experts from other field of study, which raises many questions including the subjectivity of experts, their availability and their self motives. So automating this process can save a lot of time and remove subjective and also save other resources. The main contributions of this research are:

\begin{itemize}
    \item Development of algorithms to extract specific data from text documents
    \item Development of word similarity measures for semantic analysis
    \item Visualization of semantic similarity that assist domain experts 
\end{itemize}

In the rest of the paper, we first present the case that no such work has been done neither by HCI community, nor by AI researchers. We then explain the methods that we have developed to extract the text for comparison purposes from documents; development of algorithms for computing word similarity and the usefulness of visualization tools. Finally, we will discuss the current shortcomings of this research; and what are our future endeavours to improve the results.

\section{Related work}
In general, there is an extensive literature on the extracting data from the document and semantic similarity between the words\cite{li2006sentence},\cite{hatzivassiloglou1999detecting}, but there are no researches that aim to solve the problem of automating the process of transfer students. This section reviews some related work and discusses the strengths and shortcomings of these methods.
\subsection{Knowledge Extraction}
The extraction process focuses on segmentation process which involves dividing the document into blocks that are the smallest logical entity and then proceed with extraction in the later stages \cite{hassan2005}. The segmentation process is divided into sub parts, which include generating neighbourhood graph, creating page divisions and generating whitespace density graph. For our research, we need to detect headings to make sure the algorithm knows from which part of the document to extract the text.  To make sure that the extracted text does not have document header/footer or text from another columns(if document has multiple columns), detection of header/footer text and multiple columns is also required. So, the most efficient way for our application is to detect the relevant headings and then analyze the area of the document which contains that heading before extracting the text. The authors have not seen such work before.

\subsection{Semantic Similarity between words}
The recent work in the area of Natural Language Processing has contributed valuable solutions to calculate the semantic similarity between words. This section reviews some related work in order to investigate the strengths and limitations of previous methods, and to identify the particular difficulties in computing semantic similarity. Related works can roughly be classified into two major categories: word co-occurrence methods and similarity based on a lexical database.

The word co-occurrence methods are often known as the ``bag of words'' method. They are commonly used in Information Retrieval (IR) systems \cite{Meadow:1992:TIR:531524}. This method has word list of meaningful words and every query is considered as a document. A vector is formed for the query and for documents. The relevant documents are retrieved based on the similarity between query vector and document vector \cite{li2006sentence}. This method has obvious drawbacks such as:
\begin{itemize}
    \item It ignores the word order of the sentence.
    \item It does not take into account the meaning of the word in the context of the sentence.
\end{itemize} 
Hence, this method cannot notice the similarity between sentences which share different words with similar meanings \cite{li2006sentence}.

The second method, similarity based on lexical database, uses a predefined word hierarchy which has words, meaning and relationship with other words which are stored in a tree-like structure \cite{li2006sentence}. While comparing two words, it takes into account the path distance between the words as well as the depth of the \textit{subsumer} in the hierarchy. The subsumer refers to the relative root node with respect to the two words in comparison. It also uses a word corpus, to calculate the \lq information content\rq of the word which influences the final similarity.
This method has following limitations:
\begin{itemize}
    \item The appropriate meaning of the word is not considered while calculating the similarity, rather it takes the best matching pair even if the meaning of the word is totally different in sentence.
    \item The information content of the word, form a corpus, differs from corpus to corpus. Hence, final result differs for every corpus.
\end{itemize}
Overall, the mentioned methods ignore the semantic information of the word. Our proposed algorithm makes use of algorithm described by Yuhua Li, David McLean, Zuhair A Bandar, James DO'shea, and Keeley Crockett \cite{li2006sentence}. and enhances it by initially identifying the meaning of the word in the context of the sentence. It also take into account the frequency of the word in lexical database to mimic the human observational similarity.
\section{Methodology}
The overall flow of the application is represented in Figure 2. The data source used here is the Course Outlines from different universities and colleges. 

\begin{figure}[htbp]
\begin{center}
\includegraphics[width=0.4\textwidth, scale =0.5]{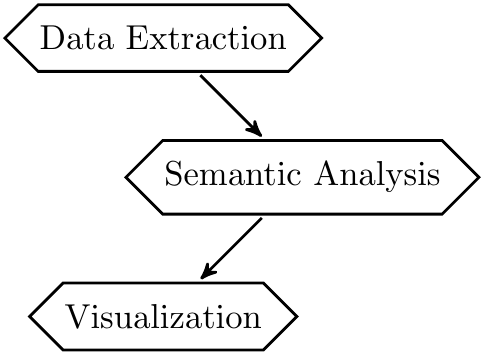}\\
\caption{Flow of the application and general modules}
\label{PdfPipeline}
\end{center}
\end{figure}

\subsection{Data Extraction}
Learning objectives are extracted first from the two documents under review. The process begins with converting the pdf document into html using the ``pdf2txt" package. It is a PDFMiner wrapper available in Python\cite{pdfminer}. The source code from HTML file is then used to extract information regarding how text is formatted using regular expressions.
For instance, the HTML code has text divided into a series of tags.\\\\

<span style="font-family: Arial-BoldMT; font-size:19px">Healthcare Information Systems</span>\\
Using the regular expression, we get the text and its corresponding text size.\\
Text: Healthcare Information Systems\\
Font Size: 19 px\\

\begin{figure}[htbp]
\begin{center}
\includegraphics[width=0.3\textwidth]{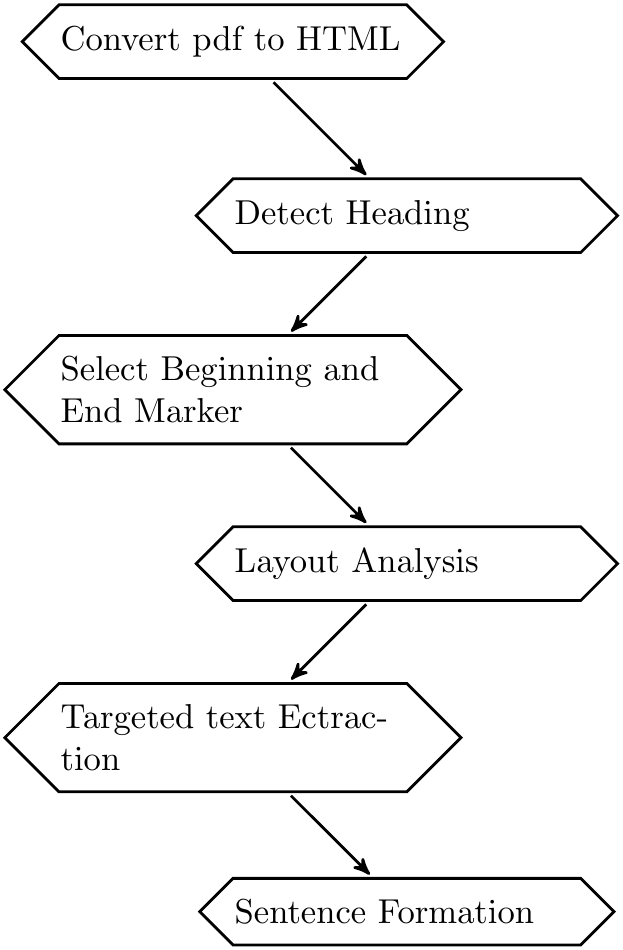}\\
\caption{Data extraction process}
\label{Fig 3}
\end{center}
\end{figure}

\subsubsection{Step 1: Heading Detection}
Using the information extracted previously, the algorithm identifies the headings in the document. Once the list of heading has been compiled the algorithm uses a keyword based approach to search for relevant headings and looks for a set of keywords in the headings to see which headings might have learning objectives listed under them. The score is calculated using the following Eq.(1):

\begin{equation}
Score = \dfrac{Number\,of\,relevant\,keyword\,in\,the heading}{Total\,number\,of\,words\,in\,the\,heading}
\end{equation}

\begin{figure}[htbp]
\begin{center}
\includegraphics[width=0.4\textwidth, scale =0.5]{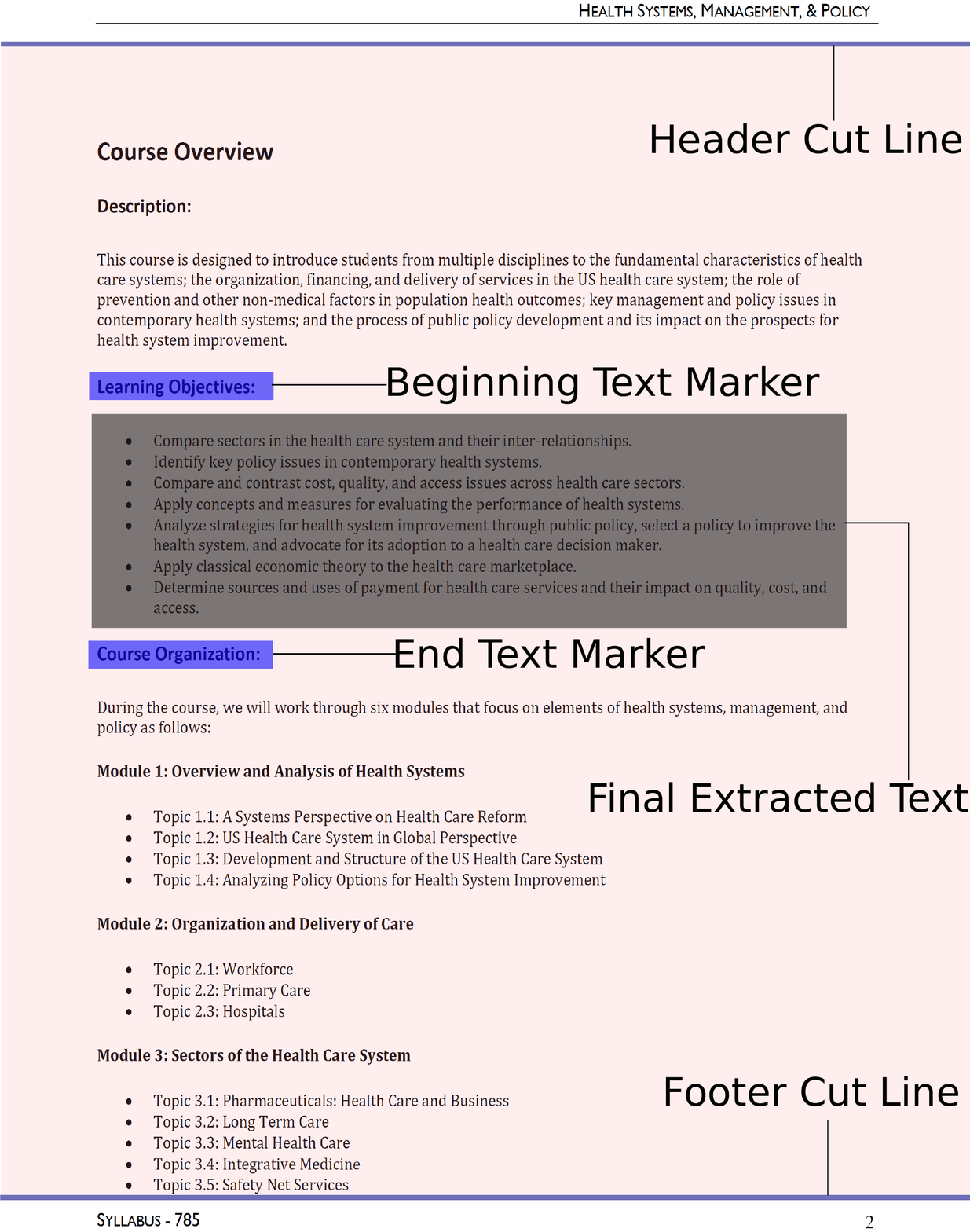}\\
\caption{Data extraction process}
\label{Fig 4}
\end{center}
\end{figure}

\subsubsection{Step 2: Looking for Beginning and End Marker}
The heading(s) with scores above the threshold are chosen as relevant. In case of multiple relevant heading, the algorithm outputs the learning outcomes considering all those paragraphs as relevant. The next step after identifying the heading is to get the text that comes under the selected heading, for that the algorithm uses the selected heading and the heading that follows the selected heading as a pair of two text markers which defines the beginning and end of the relevant text.
\subsubsection{Step 3: Layout Analysis}
Once the algorithm has located the two headings, it analyses the layout of the document by detecting white spaces and uses that to detect header/footer width and number of columns the text is divided in. 
\subsubsection{Step 4: Targeted Text Extraction}
After the layout analysis is done the information is used to extract text from a specific region of the document which ensures extraction from one column in case of multiple columns and also ensures that header/footer text don't get included in the extracted text. The extraction from specific region of the document is made possible by using the \lq PDFBox\rq  library available in Java\cite{pdfbox2014}. The algorithm then searches for two headings that it found earlier in the extracted text to get the text in between the two headings.
\subsubsection{Step 5: Sentence Formation}
The output text from the Step 4 is in the form of a continuous string and needs to be divided into sentences. To achieve that the algorithm analyses the output text for period symbol and bullet points, which help it to mark the beginning and end of sentences and generate a list of learning outcomes from the document.

\subsection{Method to compute semantic similarity between two words}
\textit{Computing Semantic similarity between two words:} The proposed method uses the lexical database,\textit{WordNet},for English language\cite{miller1995}, from the Princeton University. We also use the Bloom's taxonomy hierarchical structure to determine the complexity level of the learning objectives \cite{krathwohl2002} which are used to calculate the sentence similarity. The steps involved in computing the word similarity are:
\subsubsection{Step 1 :Identifying the words to be compared} Before calculating the semantic similarity between words, it is important to identify the words for comparison. So, we use word tokenizer and \lq part of speech tagging technique\rq as implemented in natural language processing toolkit, NLTK \cite{bird2009natural}. This step filters the input sentence and tags the important words into their  ``part of speech" and labels them accordingly.
\subsubsection{Step 2: Identifying sense of the word}
Next, we identify the appropriate sense of the word in the context of the statement. This is done by disambiguation function of Pywsd, a NLTK based python library \cite{tan2014pywsd}. This function returns appropriate sense (synset) of the word in the context with WordNet. 
\subsubsection{Step 3: Calculating shortest path distance}
In WordNet, words are arranged in sets of synonyms called ``synset''. In order to compare two synsets, it is important to take into account the length between two synsets and hierarchical position of the synsets. As the length between senses increases, the similarity decreases. Taking this into account, a previously established function is used \cite{li2006sentence}:
\begin{equation}
f(l)=e^{-\alpha l}
\end{equation}
where is $\alpha$ a constant and $\alpha \in [0,1]$.

\subsubsection{Step 4: Calculating hierarchical level information}
Also, words at upper layer of the hierarchy have less similarity than the words at the lower level of hierarchy. This is the effect of generalization as we move up the hierarchy. So, we need to scale up if the words that are being compared subsume the words at lower level and scale down if they subsume words at higher lever. Taking this behaviour into account, $g(h)$ is a monotonically decreasing function with respect to the depth $h$ \cite{li2006sentence}.
\begin{equation}
g(h)=\dfrac{e^{\beta h}-e^{-\beta h}}{e^{\beta h}+e^{-\beta h}}
\end{equation}
where $\beta $is a constant and $
\beta \in (0,1].\\
$
For WordNet, the optimal values of $ \alpha $ $ \beta $ are 0.2 and 0.45 respectively as reported previously \cite{li2003approach}.
\subsubsection{Step 5: Combining the word frequency}
Along with statistical information from WordNet, we have also taken into account the information content of the word so that the result could be as close as human observational similarity. We weight the significance of a word using its information content \cite{resnik1995using}. To mimic this behavior, we have taken into account the frequency of the senses of the word from the wordnet. The sense with maximum frequency is compared with its counter part from the other word. The similarity between these senses, is calculated using $ f(l)$ and $ g(h)$ \cite{li2006sentence}:
\begin{equation}
s(w1,w2)=f(l).g(h)=e^{-\alpha l}.\dfrac{e^{\beta h}-e^{-\beta h}}{e^{\beta h}+e^{-\beta h}}
\end{equation}
\begin{table}
  \centering
  \begin{tabular}{l r r r}
    {\small\textit{Name}}
    & {\small \textit{Synset}}\\
    \midrule
    Able & Synset(\lq able.a.01'))  \\
    design & Synset(\lq design.v.04')) \\
    implement & Synset(\lq implement.v.01')) \\
    basic & Synset(\lq basic.a.01')) \\
    programming& Synset(\lq program.v.02'))\\
    including&Synset(\lq include.v.03')), \\
    statements& Synset(\lq statement.n.01')),\\
    control& Synset(\lq control.n.11')),\\
    structures& Synset(\lq structure.n.04'))\\
    methods& Synset(\lq method.n.01'))\\
  \end{tabular}
  \caption{Words and corresponding synsets for \textit{s1}}~\label{tab:table1}
\end{table}
For \textit{s2}, the list of tagged words is in Table 2.
\begin{table}
  \centering
  \begin{tabular}{l r r r}
    {\small\textit{Name}}
    & {\small \textit{Synset}}\\
    \midrule
    Construct &Synset(\lq concept.n.01'))  \\
    computer & Synset(\lq computer.n.01')) \\
    algorithms& Synset(\lq algorithm.n.01'))\\
    solve&Synset(\lq solve.v.01')) \\
    simple&Synset(\lq simple.a.01')) \\
    problems&Synset(\lq problem.n.02'))\\
    basic& Synset(\lq basic.a.01'))\\
    computer& Synset(\lq computer.n.01'))\\
    programs& Synset(\lq program.n.08'))\\
    using& Synset(\lq use.v.03'))\\
    formal& Synset(\lq formal.a.03'))\\
    syntax&Synset(\lq syntax.n.03'))\\
    programming&Synset(\lq scheduling.n.01'))\\
    language&Synset(\lq language.n.01'))\\
  \end{tabular}
  \caption{Words and corresponding synsets for \textit{s2}}~\label{tab:table2}
\end{table}

\subsubsection{Step 6: Calculating the similarity}
Finally, for two words $w1$ and $w2$, average of the similarities calculated in \textit{Step 5}, similarity corresponding to the disambiguated word pair and similarity corresponding to the highest frequency word pair, gives the final similarity between two words. To elaborate the procedure further, please see  a detailed example below:\\
Input statements: \\
$s1$={Able to design and implement basic programming solutions including statements, control structures, and methods.}

$s2$={Construct fundamental computer algorithms to solve simple problems. Create basic computer programs using the formal syntax from a high-level, object-oriented programming language.}
\begin{figure*}
  \centering
  \includegraphics[width=1.75\columnwidth]{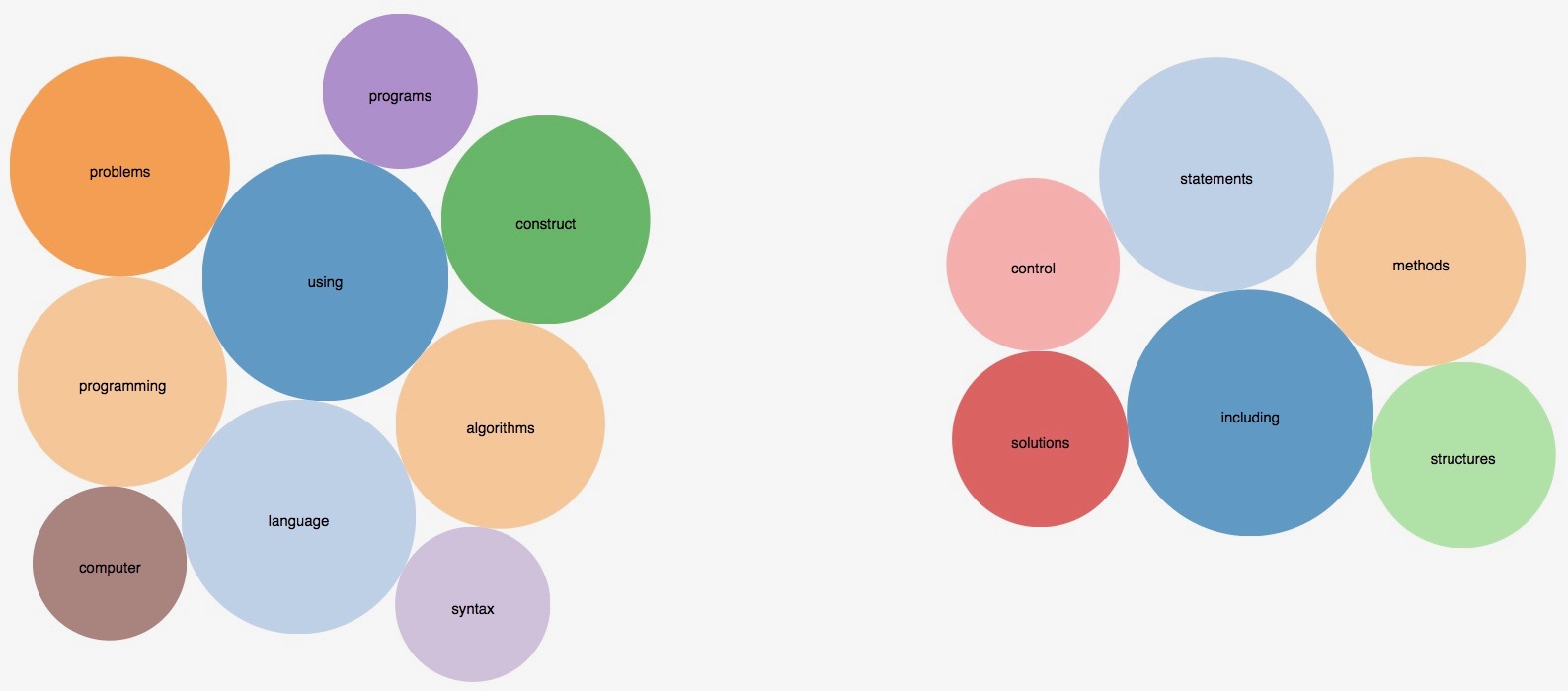}\\
  \textbf{Figure 5. Word cloud for statement 1 and statement 2 depicting weight of the word with respect to other statement}
\end{figure*}
\begin{figure*}
  \centering
  \includegraphics[width=1.75\columnwidth]{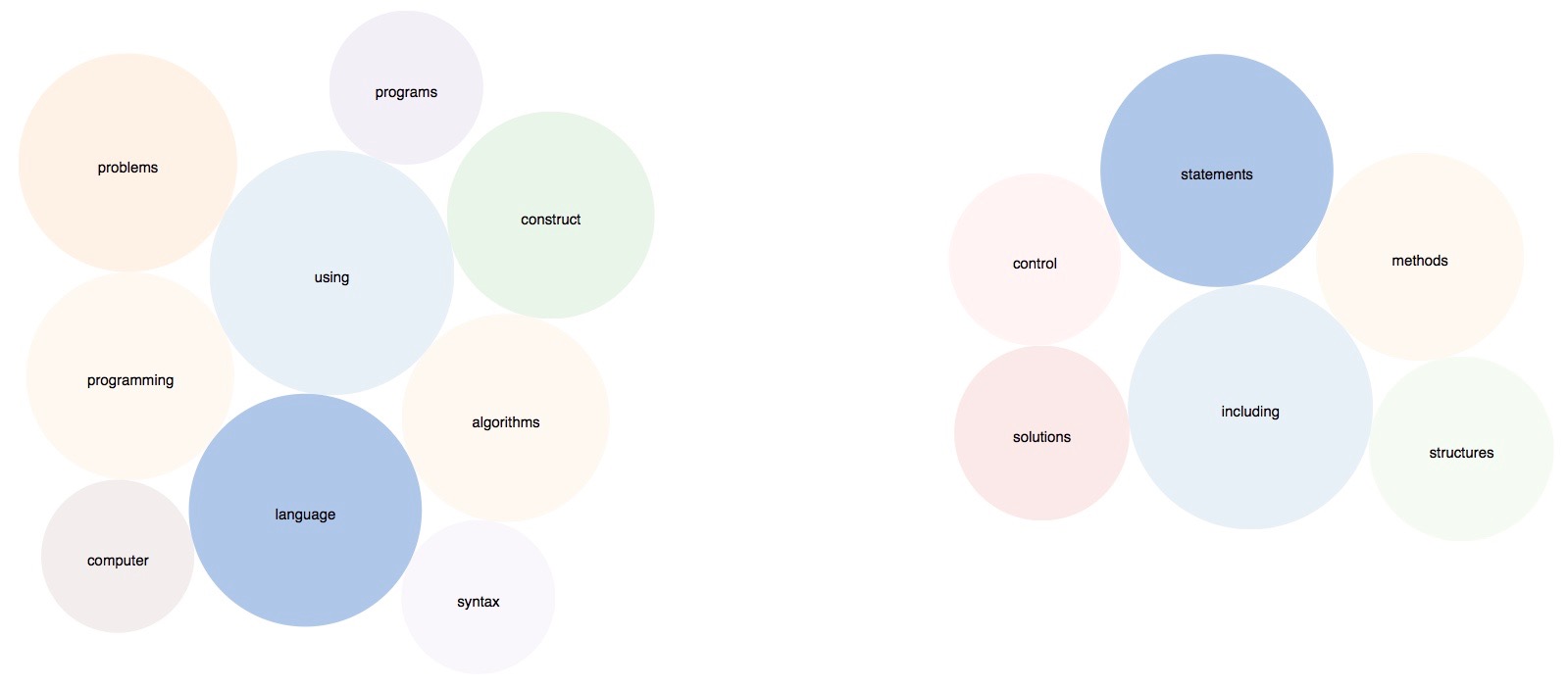}\\
  \textbf{Figure 6. Word from sentence 1 and similar word from sentence 2 having one to one relationship}
\end{figure*}
\begin{figure*}
  \centering
  \includegraphics[width=1.75\columnwidth]{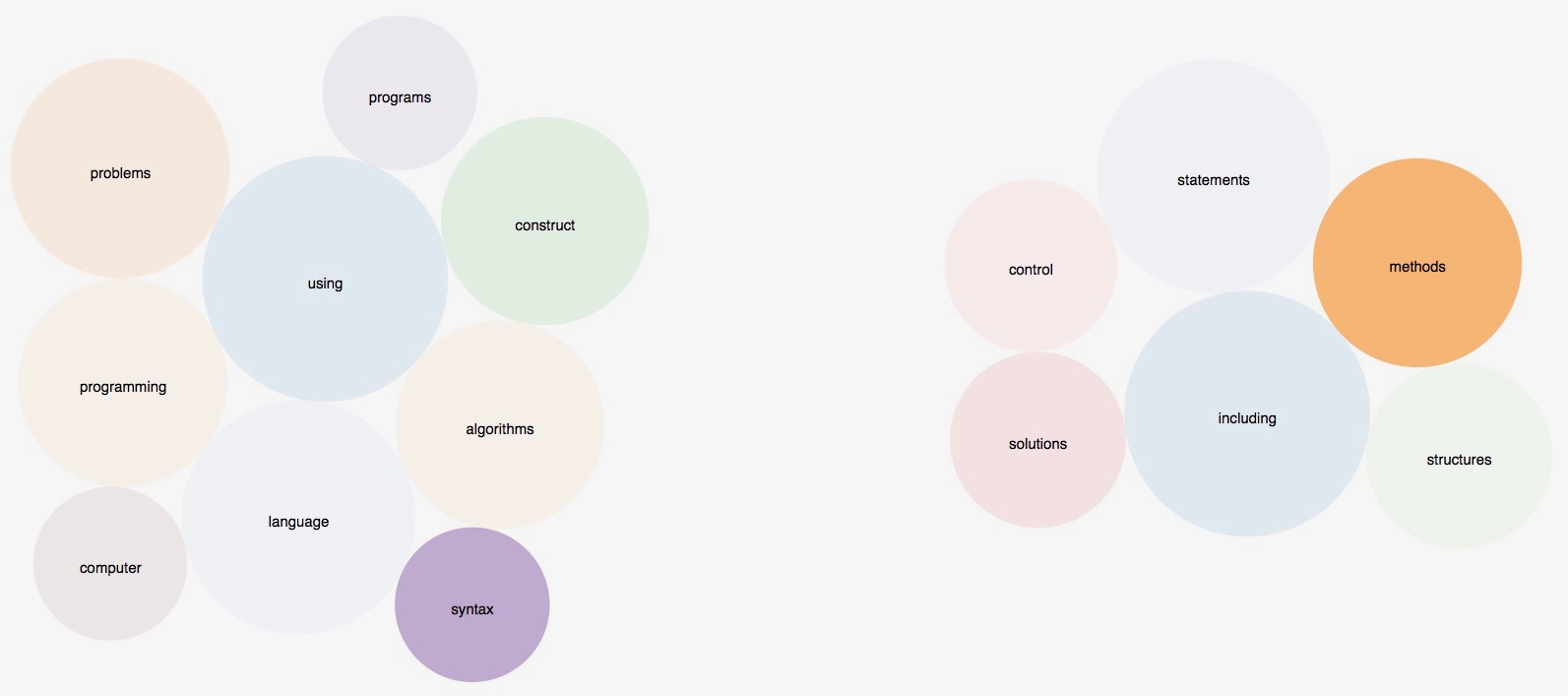}\\
  \textbf{Figure 7. Word from sentence 1 and similar word from sentence 2 having many to one relationship}
\end{figure*}
\begin{figure*}
  \centering
  \includegraphics[width=0.95\columnwidth]{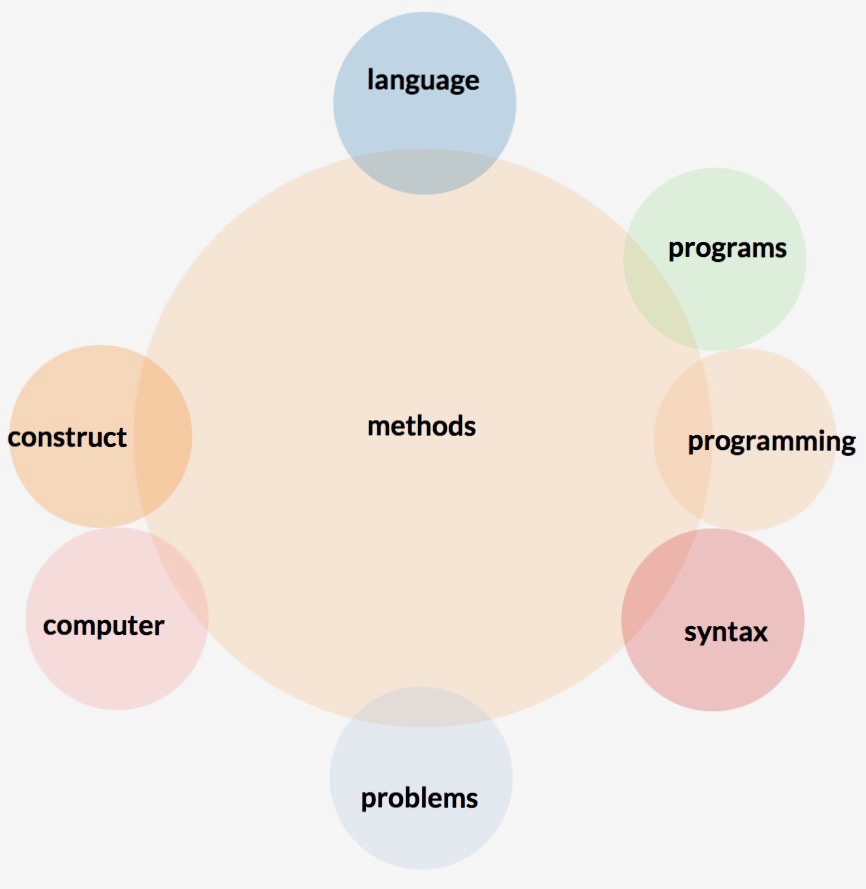}\\
  \textbf{Figure 8. Venn diagram of similarity between a word from sentence 1 with the words from sentence 2 }
\end{figure*}\\
After executing \textit{Step 1} and \textit{Step 2}, we get following list of words and corresponding senses.
For \textit{s1}, the list of tagged words is in Table 1.

\textit{Step 3:} Calculate the shortest path length between the synsets. As an example, consider following two words from \textit{s1} and \textit{s2}.\\
``solutions construct''=0.110803158362\\ \\
\textit{Step 4:} Calculate the hierarchical distance between two synsets.\\
``solution construct''=0.99939310594\\ \\
\textit{Step 5:} The similarity for disambiguated words pair is:\\
``solution construct''=0.245692969585 \hspace{28mm} (4)\\
The similarity for the `most frequent sense' word' pair is:\\
``solution construct''=0.110735912584 \hspace{28mm} (5)\\\\
\textit{Step 6:} Average of (4) and (5) gives the final similarity.
Hence,\\
\textit{similarity(solution,construct)}=0.178214441\\
This process is repeated for every keyword from sentence 1 against every keyword from sentence 2. These similarities will further be used to calculate the similarity between whole sentences. 

\subsection{Method to visualize similarity indexes}
The visualization of similarity between two words has been implemented in d3 javascript library \cite{bostock2011d3}. The power of d3 for producing dynamic, interactive visualizations in web browsers influenced our decision of tool to be used for visualization.

The primary idea of this visualization is to be able to understand the semantic similarity between the all the keywords of the two sentences being compared. Visualizations is broadly divided into three parts: a) keywords cloud b) relationship between keywords c) semantic similarity between keywords
\subsubsection{a) Keywords cloud}
The keywords cloud is created separately for both the sentences, containing keywords from respective sentences. The \textit{d3 physics} module is responsible for creating ``bubbles'' which represents keywords and the radius of the bubble depends on its similarity index calculated in the methodology to compute semantic similarity between two words. The radius of each bubble is relative. The word with maximum similarity index has a bubble with largest radius. With respect to the largest bubble, the radius of other bubbles decreases according to their similarity index. Figure 5 represents the generated keyword clouds for both the sentences.

\subsubsection{b) Relationship between words}After establishing the word cloud, the next step is to identify the relationship between words. This relationship signifies the word from other word cloud for which the similarity index holds. Each word would be connected to word from other word cloud. The relationship could be one to one or one to many. Figure 6 represents one to one relationship and Figure 7 represents one to many relationship.

\subsubsection{c) Semantic similarity between keywords}
To get the overall idea of similarity of the word with respect to words from other sentence, we used Venn diagrams as represented in Figure 8. The word in the center is the word from first sentence and other words surrounding it are the words from second sentence. The overlap between bubbles conveys the similarity. 



\section{Discussion}
This paper presents our ongoing research to automate the process of deciding the similarity between two courses. The base criteria for this decision is the learning outcomes(LO) of two courses. The process starts with extraction of the text for LOs. Since there is no standard followed to design the structure of the course outline, it is a challenging task to find and extract the LO. There is no algorithm currently available to extract the LO. The proposed algorithm, takes into account the text format,  document layout and locates the learning outcomes. We have tested our algorithm on 86 documents from various universities and institutions. The algorithm was able to tag the LOs from 78 documents correctly. The percentage accuracy is 90.69. The rest of the documents are far too generic and some had no identifier to detect the LO. The next step in improving the data extraction algorithm would be to make it compatible with more complex document layouts. To achieve this we will need to enhance the algorithm to analyze white spaces.
The extracted LOs are then fed into the next part of the system to calculate the semantic similarity between LOs in order to decide if two course are similar and whether they are eligible for credit transfer of the student. Disambiguation is an important part of the modified algorithm. Instead of starting with word similarity head on, we find the meaning of the word from WordNet and use this information. We tested this algorithm on the theme word pairs from Herbert Rubenstein and John B Goodenough's \lq Contextual correlates of synonymy\rq \cite{rubenstein1965contextual}. The results are better for the words having similar meaning. For other cases, the results are close to the algorithm implemented in \cite{li2006sentence}. Figure 8 represents the comparison of the results. The current algorithm uses the general purpose English language lexical database. Hence results are not accurate with respect to the education domain. The future work in this area of research includes - integrating the Bloom's taxonomy and educational anthologies to get specific results. 
\section{Conclusion}
In this research, we solved the problem of accurately tagging and extracting the learning outcomes(LOs) from course outlines. We also achieved better word similarity index in the context of semantic similarity. 
The results suggest that our algorithms and visualizations are helpful in supporting the project managers who are assigned the tasks of preparing transfer programs.

\section{Acknowledgments}
We would like to acknowledge the financial support provided by ONCAT(Ontario Council on Articulation and Transfer) through Project Number- 2017-17-LU,without their support this research would have not been possible. We are also grateful to Salimur Choudhury for his insight on all aspects of this project, Fahad Wali for reviewing and proofreading the paper.

%
%
%
%
%
\balance{}

\bibliographystyle{SIGCHI-Reference-Format}
\bibliography{sample}

\end{document}